\documentclass[]{article}
\usepackage{amsmath,amsthm,amscd,amssymb}
\theoremstyle{plain}
\usepackage{epsf}
\usepackage{epsfig}
\newlength{\mycolspace}
\theoremstyle{definition}

\theoremstyle{remark}

\newcommand{\calR}{{\cal{R}}}

\newcommand{\calT}{{\cal{T}}}

\newcommand{\calV}{{\cal{V}}}

\newcommand{\X}{{\bf{X}}}
\newcommand{\E}{{\bf{E}}}
\newcommand{\J}{{\bf{J}}}
\newcommand{\B}{{\bf{B}}}
\newcommand{\V}{{\bf{V}}}
\newcommand{\bO}{{\mbox{\boldmath $0$}}}
\newcommand{\n}{{\bf{n}}}

\newcommand{\w}{{\bf{w}}}
\newcommand{\z}{{\bf{z}}}
\newcommand{\bphi}{{\mbox{\boldmath $\phi$}}}
\newcommand{\brho}{{\mbox{\boldmath $\rho$}}}

\newcommand{\fr}{{^F\hspace{-.02in}R}}

\newcommand{\tr}{\operatorname{tr}}

\begin{document}
\title{Dynamics of Relativistic Flows}
\author{C. Chicone\\Department of Mathematics\\University of
Missouri-Columbia\\Columbia, Missouri 65211, USA
\and B. Mashhoon\thanks{Corresponding author. E-mail:
mashhoonb@missouri.edu (B. Mashhoon).} \\Department of Physics and
Astronomy\\University of Missouri-Columbia\\Columbia, Missouri 65211, USA
\and B. Punsly\\
4014 Emerald Street, No. 116\\
Torrance, California 90503, USA}
\maketitle
\begin{abstract}
Dynamics of relativistic outflows along the rotation axis of a Kerr
black hole is investigated using a simple model that takes into account
the relativistic tidal force of the central source as well as the Lorentz force
due to the large-scale electromagnetic field which is assumed to be
present in the ambient medium. The evolution of the speed
of the flow relative to the ambient medium is studied. In the force-free
case, the resulting equation of motion predicts rapid deceleration
of the initial flow and an asymptotic relative speed with a Lorentz
factor of $\sqrt{2}$. In the presence of the Lorentz force, 
the long-term relative speed of the clump
tends to the ambient electrical drift speed.
\end{abstract}
\section{Introduction}
\begin{figure}
\centerline{\psfig{file=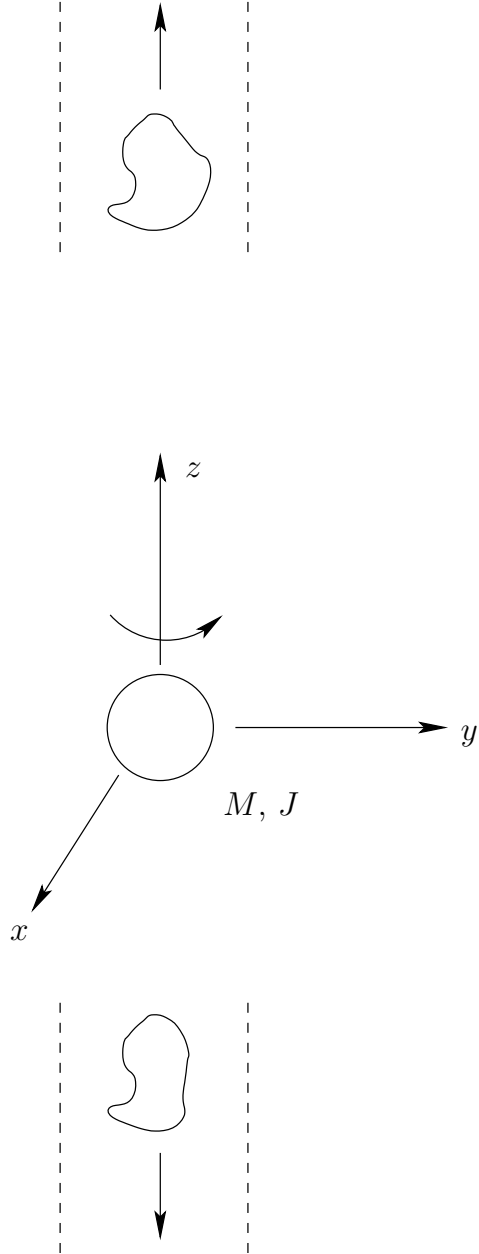, width=15pc}}
\caption{Schematic diagram representing jet clumps moving relative
to the
ambient medium along the rotation axis of a central black hole.
\label{fig:1}}
\end{figure}

A significant feature associated with quasars and active galactic nuclei is the existence
of relativistic outflows known as ``jets.'' The jets frequently come in pairs
that originate from a central ``engine'' that is conjectured to be a massive 
rotating black hole surrounded by an accretion disk. 
Rather similar phenomena are also observed in most of the X-ray binary
systems in our galaxy; therefore, these are usually called ``microquasars.''
It is thought that the jets are emitted along the rotation axis of the Kerr
black hole. 
The double-jet structure is consistent with the symmetry of a Kerr
source under reflection about its equatorial plane as illustrated
in Figure~\ref{fig:1}. The observational aspects of the motion of astrophysical jets
are discussed in~\cite{ref1,ref2,fen,fen2}.

The purpose of this paper is to discuss a relativistic tidal effect
that can be used to predict the  speeds of astrophysical jets relative
to their ambient media. 
The theory of the formation and propagation of 
jets per se is beyond the scope of this paper.  
We will consider a Kerr black hole with plasma outflow along its 
rotation axis and study the dynamics of a clump in the flow 
\emph{after it has emerged from its source environment}.
To explore the transient role of the novel relativistic tidal
effect near the source, we assume that large-scale electromagnetic
fields are dynamically important. We thus ignore the 
complicated plasma processes that actually determine the electrical drift
velocity of the ambient medium. Our approach may be contrasted with
the purely hydrodynamic models of relativistic outflows: all
force-free as well as MHD wind models assume at the outset that 
the plasma flows at the drift velocity. 
It is important to emphasize 
the idealized nature of the model employed here; our purpose  is 
simply to illustrate the regime in which previously ignored relativistic tidal
forces are significant for the dynamics of relativistic outflows.

The mechanism 
for clump formation and the possible values of the clump's initial speed 
as well as other characteristics are beyond the scope of our paper.
Once emitted, the motion of the clump relative to the ambient medium is subject to the ubiquitous tidal forces of the
central source together with the Lorentz force due to  the 
large-scale electromagnetic
field that is assumed to be present in the ambient medium. 
In addition to various
other simplifying assumptions throughout this work, 
all radiative as well as nonlinear plasma
effects~\cite{BLO,AWG} are also neglected here. 

With respect to a background Fermi frame that is fixed on the ambient
medium with coordinates $(T,\X)$
such that $\X=(X,Y,Z)$ and the $Z$-axis represents the axis of rotation
of the black hole,
the equation of motion of a plasma clump along the rotation axis relative
to the ambient medium is given by
a ``Newtonian'' equation of motion, involving the superposition of
external accelerations, of the form
\begin{equation}\label{1.1}
\frac{d^2Z}{dT^2}=A_T+A_L,
\end{equation}
where $A_T$ is the relativistic tidal acceleration due to
the central rotating mass and $A_L$ is the Lorentz acceleration
due to the electromagnetic field of the ambient
medium.
The nature of these accelerations are discussed 
in turn in Sections~\ref{sec:2}, \ref{sec:3} and~\ref{sec:new4}.  The main
dynamical features of the resulting equation of motion are described in 
Section~\ref{sec:4}.
 Section~\ref{sec:5} contains a brief discussion of our results.
We employ units such that $c=1$.

\section{The Generalized Jacobi Equation}\label{sec:2}
Imagine a background global inertial system with coordinates 
$(T ,\X )$ and two neighboring test particles in free fall in a 
Newtonian gravitational potential $\Phi (\X )$. Let $\X_{1}(T)$ 
and $\X_{2}(T)$ denote the paths of the two particles. Neglecting 
the gravitational attraction between the test particles, their 
relative position $\X (T) =\X_{1}(T)-\X_{2}(T)$ is 
determined by the tidal equation
\begin{equation}\label{2.1} 
     \frac{d^{2}X^{i}}{dT ^{2}} +K_{ij}(T )X^{j}=0,
    \end{equation}
    where $K=(K_{ij}¥)$ is a symmetric matrix given by
    \begin{equation}\label{2.2}
	K_{ij}=\frac{\partial ^{2}\Phi}{\partial X^{i}\partial 
	X^{j}}
    \end{equation}
    evaluated along the orbit of one of the particles designated as 
    the reference trajectory. Note that $\tr (K)=\nabla ^{2}\Phi 
    =4\pi G\rho_{N}$, where $\rho_{N}$ is the Newtonian mass 
    density of the source. If the motion occurs exterior to the 
    source, then the matrix $K$ given by \eqref{2.2} is trace-free 
    and harmonic. The tidal acceleration in \eqref{2.1} has been 
    evaluated to linear order in the separation of the particles.
    
    In the theoretical framework of general relativity, the free 
    neighboring test particles follow timelike geodesics of the 
    spacetime manifold. Their relative motion is described by the 
    geodesic deviation equation. This equation may be expressed in 
    terms of a Fermi normal coordinate system that is constructed 
    along the reference geodesic. A Fermi frame is an almost inertial 
    coordinate system in a finite cylindrical region along the 
    worldline of the reference geodesic~\cite{s3}. 
We now let $(T,\X)$ refer to 
    the Fermi coordinates in such a system, where the spatial 
    origin $\X =0$ is occupied by the fiducial test particle. Then, 
    the Jacobi equation \eqref{2.1} describes the relative motion of 
    the other test particle in Fermi coordinates and
    \begin{equation}\label{2.3}
	K_{ij}=\fr_{0i0j},
	\end{equation}
	provided that the speed of relative motion is negligibly small 
	compared to $c=1$. Here
	\begin{equation}\label{2.4} 
	    \fr_{\alpha\beta\gamma\delta}=R_{\mu\nu\rho \sigma}\lambda 	    
	    ^{\mu}_{(\alpha)} \lambda^{\nu}_{(\beta)} \lambda^{\rho}_{(\gamma)} 
	    \lambda^{\sigma}_{(\delta )},
	    \end{equation}
so that the Riemann tensor in the Fermi frame evaluated along the 
reference trajectory $(T,\bO)$ is in effect the Riemann tensor 
projected onto the nonrotating orthonormal tetrad 
$\lambda^{\mu}_{(\alpha)}$ that is carried along the fiducial 
geodesic and upon which the Fermi system is constructed. 
Specifically, the temporal axis of the fiducial observer 
$\lambda^{\mu}_{(0)}=dx^{\mu}/d\tau$ is the vector tangent to 
its worldline, $\tau$ is its proper time and 
$\lambda^{\mu}_{(i)},i=1,2,3$, are its spatial axes. One can show 
explicitly that in the Newtonian approximation of general relativity 
\eqref{2.3} reduces to \eqref{2.2}, so that the formal analogy 
developed here has a deep physical basis.

The deviation between geodesics is taken into account only to linear 
order in the Jacobi equation; in fact, higher-order terms can be 
neglected so long as the deviation is very small compared to the 
radius of curvature of spacetime $\calR$. On the other hand, in 
certain circumstances, the relative speed of neighboring geodesics 
may not be small compared to unity. The corresponding generalization 
of the Jacobi equation is given by
\begin{multline}\label{2.5}
    \frac{d^{2}X^{i}}{dT^{2}}+ \fr_{0i0j}X^{j}+2\,\fr_{ikj0} 
    V^{k}X^{j}\\
    + (2\,\fr_{0kj0}V^{i}V^{k} +\frac{2}{3}\, \fr_{ikj\ell} 
    V^{k}V^{i}+\frac{2}{3}\, \fr_{0kj\ell}V^{i}V^{k}V^{\ell} 
    )X^{j}=0.
\end{multline}
This generalized Jacobi equation has been discussed in detail in 
\cite{cm}. If we limit our attention to motion along a fixed 
direction in space, say along the $Z$-direction that characterizes 
the rotation axis of a Kerr system in  the Fermi frame, then 
\eqref{2.5} reduces to
\begin{equation}\label{2.6}
    \frac{dV}{dT}+\kappa (1-2V^{2})Z=0,
    \end{equation}
    where $V=dZ/dT$ and $\kappa =\fr_{0303}$. This equation has an 
    interesting feature that was first pointed out in \cite{cm}: 
    Special solutions of \eqref{2.6} exist that correspond to uniform 
    rectilinear motion with limiting velocities of $V_T=\pm 
    (2)^{-\frac{1}{2}}$. For $\kappa <0$, these correspond to 
    attractors if
    \begin{equation}\label{2.7}
	\int^{\infty}_{T_{0}}\kappa (T) (T+C_{0})dT=-\infty
	\end{equation}
	for an arbitrary real number $C_{0}$ and an arbitrary initial time 
	$T_{0}$ \cite{cm}.
	
	It is useful to consider the application of \eqref{2.6} to motion 
	along the rotation axis of a Kerr source. The main results are 
	expected to hold qualitatively for any central rotating and 
	axisymmetric configuration such as, for instance, the Kerr spacetime 
	endowed with an infinite set of higher moments \cite{QM}. The motion 
	of free test particles along the symmetry axis of an exterior
 Kerr spacetime 
	representing the stationary 
gravitational field of a source of mass $M$ and 
	angular momentum $Ma$ is given by
	\begin{equation}\label{2.8}
	    \left(\frac{dr}{d\tau}\right)^{2}=\gamma^{2}-1+ 
	    \frac{2GMr}{r^{2} +a^{2}} 
	    \end{equation}
	    in Boyer-Lindquist coordinates. Here $\gamma>0$ is an 
	    integration constant; in fact, in the $\gamma\ge 1$ 
case it has the 
	    interpretation of the particle's Lorentz factor at 
	    infinity $(r \to \infty)$. We take the reference geodesic to 
	    represent the motion of the ambient medium with $dr/d\tau > 0$
 along the
$Z$-axis in the exterior Kerr 
	    spacetime and we construct a Fermi coordinate system $(T,X,Y,Z)$ 
	    in the neighborhood of this geodesic such that as the reference 
	    geodesic is approached, $(T,\X )\to (\tau ,\bO )$. Evaluating the 
	    Riemann curvature tensor in this frame, we find that $\kappa \to 
	    k$ is given by
	    \begin{equation}\label{2.9}
		k=-2\frac{GMr(r^{2}-3a^{2})}{(r^{2}+a^{2})^{3}}.
		\end{equation}
		That \eqref{2.9} is explicitly independent of $\gamma$ is a 
		consequence of the fact that Kerr spacetime is of type $D$ in the 
		Petrov classification and hence the axis of rotation provides two 
		special tidal directions for ingoing and outgoing trajectories 
		\cite{mm,bee}. Moreover, $a=0$ corresponds to radial motion in the 
		exterior Schwarzschild spacetime.
For a Kerr black hole $0<a\le GM$, where $a=GM$ corresponds to an
extreme Kerr black hole. 
To emphasize that our treatment 
		involves motion along the rotation axis of a Kerr system, it is 
		useful to introduce Cartesian coordinates $(x,y,z)$ defined in terms 
		of Boyer-Lindquist coordinates $(r ,\theta ,\varphi )$ as 
		usual: $x=r\sin \theta \cos \varphi$, $y=r\sin \theta \sin \varphi $ 
		and $z=r\cos \theta$. In Section~\ref{sec:4}, we will employ 
		this $z$-coordinate along the Kerr rotation axis that in the Fermi 
		system becomes the $Z$-axis.
Moreover, we will assume that
the reference particle at $Z = 0$ is such that $\gamma = 1$ and
$r > \sqrt{3}\,a$ along
its geodesic path. 
The $\gamma=1$ condition implies that the reference particle moves away
from the black hole forever, coming to rest at infinity. 
The collection of such particles provides an ambient medium with
respect to which the motion of the jet clump can be described. 
That is, this ambient medium provides a dynamical way
of characterizing the rest frame of the black hole. 
		
		Imagine two neighboring test particles at $r$ and $r+\delta r$ 
		along the rotation axis of a central source. Far away from the 
		source, the Newtonian gravitational acceleration of the reference 
		particle at $r$ is $-GM/r^{2}$ and for the other particle at 
		$r+\delta r$ is $-GM/(r+\delta r)^{2}$. The relative or tidal 
		acceleration is then given by $2GM\delta r/r^{3}$, which by 
		\eqref{2.9} is $-k\delta r$ for $r >>GM$ and $r >>a$. 
		Therefore, \eqref{2.6} and \eqref{2.9} reduce to the familiar 
		Newtonian result once the relativistic tidal acceleration 
		proportional to $V^{2}<<1$ is neglected.
		
\begin{figure}[p]
\centerline{\psfig{file=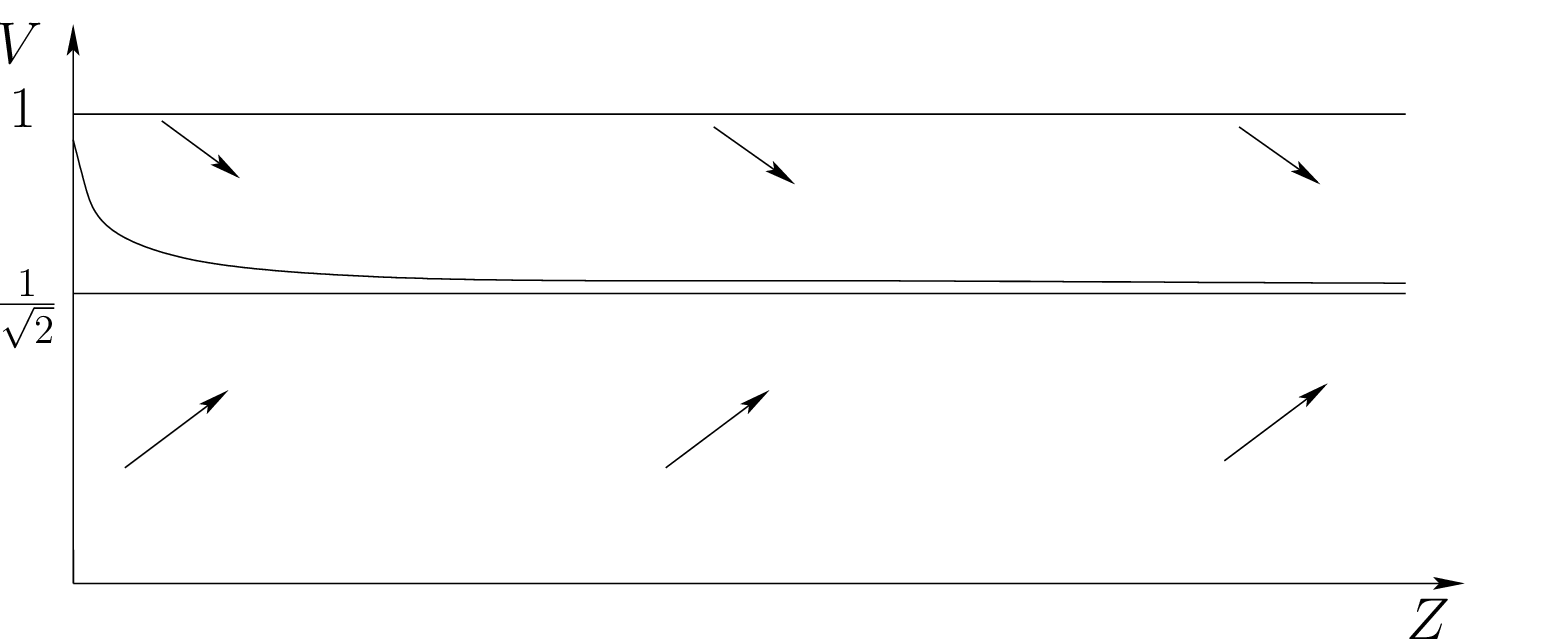,width=30pc}}
\centerline{\psfig{file=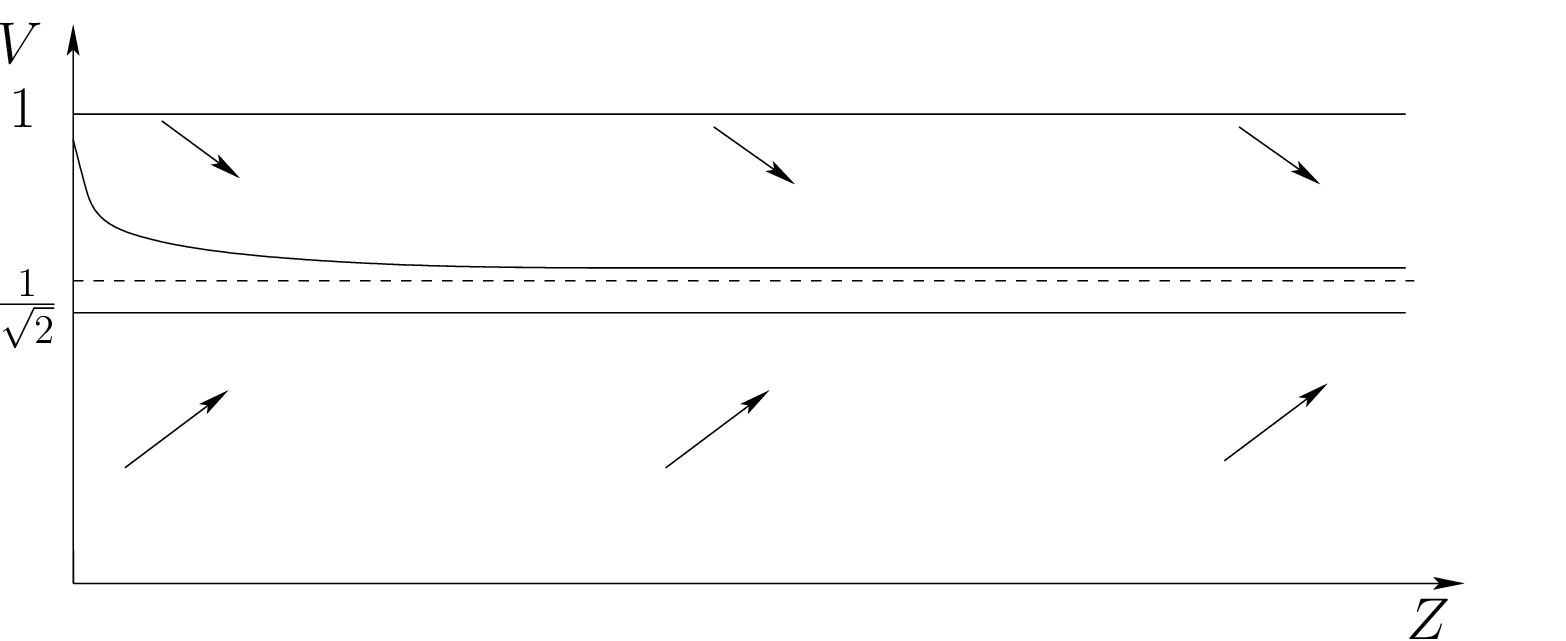,width=30pc}}
\caption{Schematic representation of direction fields and  trajectories for
the force-free case given by the generalized Jacobi equation.
Equation~\eqref{2.10} is integrated with the initial conditions
that at $T=0$, $r=r_0>\sqrt{3}\, a$, $Z=0$ and $V=V_0>0$;
thus, initially $dV/dZ=0$ as well.
The top panel depicts a trajectory for $\gamma=1$. In this
case the velocity is asymptotic to $1/\sqrt{2}$.
 The bottom panel depicts a trajectory
for $\gamma>1$. In this case, the trajectory is asymptotic to 
some constant velocity, depicted by the dashed line, 
that is closer to $1/\sqrt{2}$ than the initial velocity.
The limiting velocity can be calculated, in principle, 
from the integration of
the generalized Jacobi equation $V dV/dZ = -k ( 1 - 2 V^2) Z$, namely, 
$1 - 2 V^2 =
( 1 - 2 V_0^2 ) \exp( 4 \int_0^Z k ( T (Z' ) ) Z'\, dZ' )$. 
For $\gamma = 1$, the integral as $Z\to\infty$ is $-\infty$ in agreement 
with~\eqref{2.7} and
hence the limiting velocity is given by $V = 1/\sqrt{2}$.
\label{fig:ff}}
\end{figure}

		The speed of a clump can be measured by observing the 
		displacement of the clump in the flow relative to the ambient medium 
		\cite{ref1,ref2,fen,fen2}. 
Therefore, consider a clump along the axis of rotation 
		moving rapidly past the reference particle belonging to the ambient 
		medium. The equation of relative motion is 
 given by \eqref{1.1} 
		with
		\begin{equation} \label{2.10}
		    A_{T}¥=-k(1-2V^{2}¥)Z,
		    \end{equation}
		    where $k$ is determined via equations \eqref{2.8} and 
		    \eqref{2.9}.
Under our assumption we have that $r^{2}¥>3a^{2}¥$, and thus $k<0$. 
Also, we have assumed that $\gamma =1$ in~\eqref{2.8} 
		     for the ambient medium 
		    surrounding the black hole; therefore,
 condition \eqref{2.7} is 
		    satisfied, the two special solutions of \eqref{2.10} are 
		    indeed attractors and 
the clump speed (predicted by our model) tends to $\approx 
		    0.7$ over time~\cite{cm}. 
In~\cite{cm} a similar 
assertion was also made for $\gamma<1$, which turns
out to be erroneous in general; in any case, it is not relevant for a black hole
source with $a \le GM$. The 
qualitative dependence of the solution of the generalized Jacobi equation
upon $\gamma\ge 1$ is depicted in Figure~\ref{fig:ff};
in constructing this figure, we consider equation~\eqref{2.10} simply as a nonlinear
differential equation and ignore its possible physical limitations.

\begin{figure}[p]
\centerline{\psfig{file=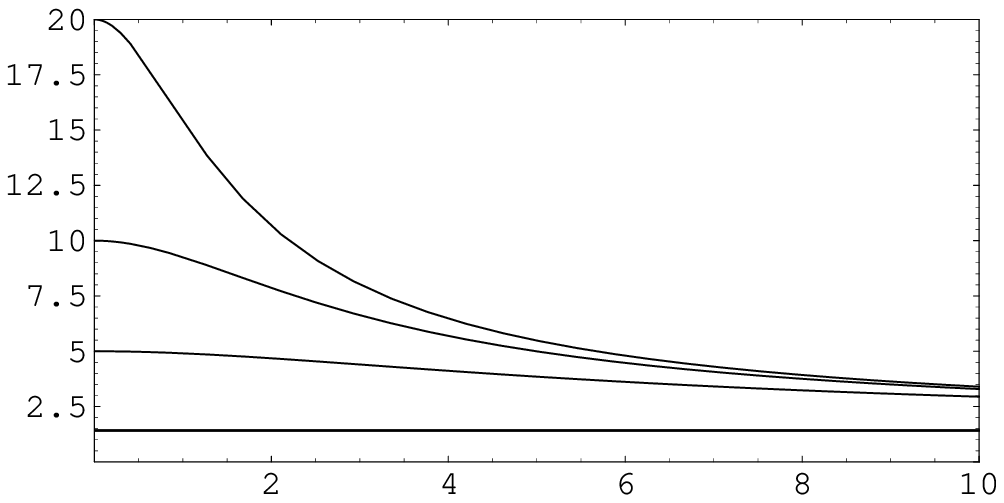,width=30pc}}
\caption{Plot of the Lorentz factor $\Gamma = ( 1 - V^2 ) ^{-1/2}$
 versus $T/(GM)$ based
on the integration of equation~\eqref{2.10} with initial data 
$r_0/(GM) = 10$, $Z(0) = 0$, $V(0)
= 2 \sqrt{6}/5 \approx 0.980$, $ 3 \sqrt{11}/10 \approx 0.991$ and 
$\sqrt{399}/20\approx 0.999$
corresponding to $\Gamma_0 = 5$, 10 and 20, 
respectively. Here we assume that $a/(GM) =1$ and $\gamma = 1$. 
At $T/(GM) = 10$ the corresponding Lorentz factors are 
$\Gamma \approx 2.945$, 3.291 and 3.398, respectively. 
For the sake of comparison, the horizontal line indicates the Lorentz factor 
$\sqrt{2}$ 
corresponding to the terminal speed.
Further away from the central source, 
the tidal forces decrease and hence the
initial deceleration is correspondingly weaker. 
For instance, with $r_0/(GM) = 100$,
the clump decelerates from $\Gamma_0 = 5$ to  $\Gamma = 3$
over a time interval given approximately by $T/(GM) = 300$.\label{fig:nf3}}
\end{figure}
It is important to recognize two salient features of
equation~\eqref{2.10} for the force-free motion of an ultrarelativistic
clump relative to the ambient medium with $\gamma = 1$. 
First, the initial deceleration of
the clump can be quite significant
\emph{within the range of validity} of the generalized Jacobi 
equation~\eqref{2.10}.  For instance, Figure~\ref{fig:nf3}
depicts the deceleration of clumps with initial 
Lorentz factors of $\Gamma_0 = 20$, 10 and 5,
where we have introduced the Lorentz factor of the clump
$\Gamma=(1-V^2)^{-1/2}$, and $\Gamma_0$ corresponds to the initial
clump speed $V_0$ at $r=r_0$ and $T=Z=0$.
Launched from $r_0 = 10\, GM$, the clumps decelerate to $\Gamma\approx 3$
in about $T = 10\, GM \approx 5\times 10^{-5}\, ( M/M_{\odot} )\, \mbox{sec}$. 
Although the distance of such a clump relative to the
reference particle increases rapidly and soon the validity of
the generalized Jacobi equation
breaks down, the subsequent motion can be measured with the same
model equation relative to the motion of 
another suitable reference particle in the ambient medium.  
This observation uses the second remarkable
feature of equation\eqref{2.10}: it does not explicitly depend on the
choice of the
reference particle in the ambient medium with $\gamma = 1$. 
To illustrate this
point, imagine a sequence of $n$ reference particles at $r_i$, 
$i = 1, 2, \cdots , n$,
all with $\gamma = 1$ along the rotation axis of the central source such 
that $r_1<r_2 <\cdots<r_i< r_{i+1}<\cdots<r_n$. 
We can set up distinct Fermi frames along the 
worldline of each of
these reference particles such that within each interval $[r_i, r_{i+1}]$ 
the generalized Jacobi equation is
valid in the Fermi frame based on $r_i$. Once the clump reaches 
$r_{i+1}$,
we can switch to the new Fermi frame based on $r_{i+1}$. 
So long as the initial
speed of the clump relative to $r_{i+1}$ is above $1/\sqrt{2}$, 
the model predicts that the clump continues to
decelerate toward the terminal speed. In this sense, the speed of the clump
\emph{relative to the ambient medium}
approaches the terminal speed over time.

In the absence of any nongravitational forces, equations~\eqref{1.1} 
and~\eqref{2.10} express the geodesic equation for the center of mass
of the clump in Fermi coordinates. Once the electromagnetic forces are
included, the general equation of motion takes the form
\begin{equation}\label{eq:12}
\rho'_m(\frac{d^2x^\mu}{ds^2}+\Gamma_{\nu\rho}^\mu 
\frac{dx^\nu}{ds}\frac{dx^\rho}{ds})=F^\mu_{\hspace{.1in}\sigma}\,j^\sigma,
\end{equation}
where $\rho'_m$ is the invariant mass density of the clump,
$F_{\mu\nu}$ is the Faraday tensor of the exterior electromagnetic
field in the ambient medium and $j^\mu$ is the charge current of the clump.
Expressed in the Fermi coordinate system $(T,\X)$, equation~\eqref{eq:12} 
reduces to equation~\eqref{1.1}, where $A_L$ is due to the Lorentz force.
In evaluating the Lorentz acceleration $A_L$ in the following section,
we employ the physically reasonable approximation that the effects of spacetime
curvature can be neglected for the sake of simplicity; thus,
we treat the Fermi frame established along the reference geodesic as an
inertial frame of reference in Section~\ref{sec:3}.

It is conceivable that the electromagnetic field configuration is such
that $F^\mu_{\hspace{.1in}\sigma}\,j^\sigma=0$ in equation~\eqref{eq:12} 
and hence the motion of
the  clump is force-free (see, for example,~\cite{JP}).
This may be the case, for instance,
beyond the launching point of the flow, a few gravitational radii from
the central source. In the
force-free case, the detailed analysis of~\cite{cm} is applicable 
in the sense explained above and the clump
speed tends to $1/\sqrt{2} \approx 0.7$, corresponding to a Lorentz
factor of $\sqrt{2} \approx 1.4$. Our results may be of interest in 
connection with Galactic superluminal sources~\cite{fen,fen2}; in fact,
as already pointed out in~\cite{cm}, some of these superluminal jets
have speeds
that may be near $1/\sqrt{2}\approx 0.7$. In the rest of this paper, we investigate
clump motion that is \emph{not} force-free.

\section{Lorentz Acceleration}\label{sec:3}
Imagine a definite clump of plasma with constant
mass $m$ moving with respect to the ambient medium with 
velocity $\V=V\widehat\n$, where $\widehat\n$ is a fixed direction in space
that will be later identified with the rotation axis of the central gravitational
source.
The Lorentz force law for the motion of the clump can be written as
\begin{equation}\label{3.1}
\frac{d}{dT}\Big(\frac{mV}{\sqrt{1-V^2}}\Big)=
  (\delta \calV)(\rho\E+\J\times\B)\cdot\widehat\n,
\end{equation}
where $\delta \calV$, $\rho$ and $\J$ are, respectively, the volume,
charge density and charge current of the clump with respect to the
inertial frame $(T,\X)$.
We assume that in the rest frame of the clump
$m=\rho_m'\delta \calV'$, where $\rho_m'$ and $\delta \calV'$
are the proper density and volume of the plasma clump,
respectively, and are assumed to be constants throughout
the motion under consideration here.
Primed quantities refer to the rest frame of the clump. 
It follows from Lorentz contraction that~\eqref{3.1} may be written as
\begin{equation}\label{3.2}
\frac{d}{dT}\Big(\frac{V}{\sqrt{1-V^2}}\Big)=\frac{\sqrt{1-V^2}}{\rho_m'}
  (\rho\E+\J\times\B)\cdot\widehat\n.
\end{equation}
Moreover, we note that for one-dimensional motion
\begin{equation}\label{3.3}
\frac{d}{dT}\Big(\frac{V}{\sqrt{1-V^2}}\Big)=(1-V^2)^{-3/2}\,\frac{dV}{dT},
\end{equation}
so that equation~\eqref{3.2} reduces to 
\begin{equation}\label{3.4}
\frac{dV}{dT}=\frac{(1-V^2)^{2}}{\rho_m'} (\rho\E+\J\times\B)\cdot\widehat\n.
\end{equation}

We now turn to the electromagnetic aspects of~\eqref{3.4} and assume
that in the rest frame of the clump $\rho'=0$ and that 
$\J'=\sigma\E'$ in accordance with Ohm's law. 
The electrical conductivity $\sigma$ is a tensor in general; however,
we take $\sigma$ to be a scalar for the sake of simplicity.
It follows from
\begin{equation}\label{3.5}
\rho'=\Gamma (\rho-\V\cdot\J)
\end{equation}
that $\rho=\V\cdot\J=VJ_\parallel$. 
This fact immediately implies that $J'_\parallel=\Gamma^{-1} J_\parallel$
using the transformation
\begin{equation}\label{3.6}
J'_\parallel=\Gamma (J_\parallel-V\rho).
\end{equation}
Thus, we have 
\begin{equation}\label{3.7}
\J'_\bot=\J_\bot,\qquad J'_\parallel=\Gamma^{-1} J_\parallel
\end{equation}
and at the same time
\begin{equation}\label{3.8}
\E'_\bot=\Gamma(\E+\V\times\B)_\bot,\qquad E'_\parallel=E_\parallel.
\end{equation}
It follows from~\eqref{3.7}, \eqref{3.8} and $\J'=\sigma \E'$
that
\begin{equation}\label{3.9}
\J=\sigma\Gamma (\E+\V\times\B).
\end{equation} 
Thus $\rho=VJ_\parallel=\sigma\Gamma V E_\parallel$ and
\begin{equation}\label{3.10}
(\rho\E+\J\times\B)\cdot \widehat{\n}=\sigma\Gamma VE^2_\parallel
+\sigma\Gamma[(\E+\J\times\B)\times\B]_\parallel.
\end{equation} 
Using the identity
\begin{equation}\label{3.11}
[(\V\times\B)\times\B]_\parallel=-V(B^2-B^2_\parallel)=-VB^2_\bot,
\end{equation}
we find that
\begin{equation}\label{3.12}
(\rho\E+\J\times\B)\cdot \widehat{\n}=\Gamma\sigma B^2
(w-\frac{B^2_\bot-E_\parallel^2}{B^2}V),
\end{equation}
where
\begin{equation}\label{3.13}
w=\frac{(\E\times\B)\cdot\widehat{\n}}{B^2}
\end{equation}
and 
\begin{equation}\label{eq:25}
\w_D=\frac{\E\times\B}{B^2}
\end{equation}
is the  electrical drift velocity~\cite{Lp}.
To proceed further, we need a model for the electromagnetic field 
configuration in the medium surrounding the black hole. This is considered
in the next section.

\section{A Simple Model}\label{sec:new4}
Let us now consider the magnetically dominated environment in which the clump
is moving. 
In some models for such an environment~\cite{BP} and in 
cylindrical coordinates $(\rho ,\phi,z)$ with $\widehat{\n}=\widehat{\z}$,
integrating Maxwell's
 equation $\nabla \cdot \B=0$ implies that $B_{z}\sim\rho^{-2}$.
 If angular momentum is approximately conserved in the electromagnetic
 field, then $B_{\phi}\sim \rho^{-1}$. Thus, at large distances from
 the source, jets and collimated winds are often considered to have 
 predominantly toroidal magnetic fields. Assuming that the plasma is
 efficient at maintaining a very small proper electric field through its
 conductive properties, this implies that $\E$ is approximately radial in
 cylindrical coordinates~\cite{BP}. 
It is clear from this brief description that in any reasonably
realistic scenario the electromagnetic field configuration would be
rather complicated. However, we seek a simple model situation
that would render the resulting equation of motion amenable to mathematical
analysis.
Therefore, we assume that
the average electric field is primarily radial,
$\E=E\,\widehat{\brho}$,
and the average magnetic field is primarily azimuthal,
$\B=B\widehat{\bphi}$,
such that $E<B$. In view of the symmetry of the central source about
its equatorial plane, in this paper we 
concentrate---for the sake of simplicity---only on the clump moving along
the positive $z$-axis. 
It follows from~\eqref{3.13} that
the electrical drift speed is given by $w=E/B<1$.
Moreover,~\eqref{3.12} takes the simple form $\Gamma(\sigma B^2)(w-V)$.

It turns out that in the rest frame of the clump
$\E'=E'\,\widehat{\brho}$ and $\B'=B'\widehat{\bphi}$,
i.e. the fields have the same configuration as in the global inertial frame
$(T,\X)$. Thus
following the analysis in \S 2.10 of~\cite{BP}, the 
conductivity $\sigma$ should be identified with $\sigma_\bot$ as the current
$\J'=\sigma_\bot\E'$
will be crossing the magnetic lines of force in the
local rest frame of the clump. 
On the other hand,
for motion along the magnetic lines of force, the electrical conductivity
$\sigma_\parallel$ would have its usual value
\begin{equation}\label{3.14}
\sigma_\parallel=\frac{n_e e^2\tau_c}{\mu_e},
\end{equation}
where $-e$, $n_e$ and $\mu_e$ are, respectively, the charge,
number density and mass of the electron in the electron-proton
plasma in the rest frame of the clump.
Here the relaxation time $\tau_c=\nu_c^{-1}$ is the average
time interval between electron collisions and hence $\nu_c$ is the average
frequency of electron collisions per second. 
It follows from the discussion in \S 2.10 of~\cite{BP} that in case of
a tenuous plasma in a strong magnetic field such that $\nu_c^2\ll\Omega_e^2$,
we have
\begin{equation}\label{3.15}
\frac{\sigma_\bot}{\sigma_\parallel}=\Big(\frac{\nu_c}{\Omega_e}\Big)^2,
\end{equation}
where $\Omega_e=eB'/\mu_e$ is the cyclotron frequency for a free electron.
Thus
\begin{equation}\label{3.16}
\frac{\sigma_\bot B^2}{\rho'_m}=\Big(\frac{\mu_e}{\mu_p}\Big)
\Big(\frac{B}{B'}\Big)^2\nu_c,
\end{equation}
where $\mu_{p}$ is the proton mass and $\rho'_m=n_p\mu_p+n_e\mu_e=n_e(\mu_p+\mu_e)\approx n_e\mu_p$,
since $\rho'=0$ implies that the density of protons in the
clump is the same as that of electrons, $n_p=n_e$, and 
$\mu_p/\mu_e\approx 1836\gg 1$.
Moreover,
\begin{equation}\label{3.17}
\B'_\bot=\Gamma(\B-\V\times \E)_\bot,
\end{equation}
which implies that 
\begin{equation}\label{3.18}
\frac{B'}{B}=\Gamma (1-wV).
\end{equation}
Combining these results with~\eqref{3.4} reveals that
the Lorentz acceleration is given by
\begin{equation}\label{3.19}
A_L=\frac{dV}{dT}=\alpha \frac{(1-V^2)^{5/2}}{(1-wV)^2} (w-V),
\end{equation}
where
\begin{equation}\label{3.20}
\alpha=\Big(\frac{\mu_e}{\mu_p}\Big)\nu_c.
\end{equation}
It is interesting to note that for copper at room temperature
$\nu_c\sim 10^{14}\,\mbox{sec}^{-1}$.

It follows from equations~\eqref{3.19} and~\eqref{3.20}
that the Lorentz acceleration $A_L$ vanishes if the
clump speed is equal to the drift speed. 
Moreover, $A_L$ is proportional to the collision frequency $\nu_c$.
This is due to the fact that only through collisions can an appropriate interior
current be established that could lead to the acceleration of the clump
along the flow direction.

It remains to provide an estimate of $\nu_c$ for the case of the electron-proton
plasma under consideration here. This is a rather complicated problem
and we therefore limit our investigation
to the simpler case of collisions
among the electrons while neglecting the motion of the protons. Thus
we assume that $\tau_c$ can be estimated using the ``self-collision time''
of electrons given by equation (5-26) of~\cite{Lp}
\begin{equation}\label{eq:32}
\tau_c^*=\frac{0.266\, \calT^{3/2}}{n_e\ln \Lambda}\, \mbox{sec},
\end{equation}
where $\calT$ is the absolute temperature and $n_e$ is the number of electrons
in the clump per $\mbox{cm}^3$. Here $\Lambda$ is essentially the ratio
of two lengths: the Debye shielding distance over the characteristic impact
parameter of electron collisions such that the deflection angle in the orbital
plane is equal to $\pi/2$. One can use Table 5.1 of~\cite{Lp} to find the
values of $\ln\Lambda$ in terms of $\calT$ and $n_e$. According to this table,
for a fixed $\calT$, $\ln\Lambda$ decreases very slowly but monotonically
as $n_e$  increases from  $n_e=1\, \mbox{cm}^{-3}$ to 
$n_e=10^{24}\, \mbox{cm}^{-3}$; for example, for $\calT=10^8\,\mbox{K}$, 
$\ln\Lambda$
monotonically decreases from $34.3$ for $n_e=1\, \mbox{cm}^{-3}$
to $6.69$ for $n_e=10^{24}\, \mbox{cm}^{-3}$. 
Thus for $\calT=10^8\,\mbox{K}$ and $n_e=10^6 \, \mbox{cm}^{-3}$, 
we find $\ln\Lambda=27.4$
and hence from~\eqref{eq:32} with $\nu_c\approx (\tau_c^*)^{-1}$ we obtain 
$\alpha\approx 0.56\times 10^{-7}\,\mbox{sec}^{-1}$. 
For fixed $\calT$, $\alpha$ increases almost linearly with $n_e$ such that
for $n_e=10^{15}\, \mbox{cm}^{-3}$ in the case under discussion, we have 
$\alpha\approx 35\,\mbox{sec}^{-1}$~\cite{Lp}.
These considerations illustrate the fact that $\alpha$ 
could have a considerable
range of values depending upon the electron 
temperature $\calT$ and density $n_e$.

\section{Equation of Motion}\label{sec:4}
The equation of  motion  of the clump is given by
\begin{equation}\label{4.1}
\frac{d^{\,2} Z}{d\,T^2}+k(T)(1-2V^2)Z
=\alpha \frac{(1-V^2)^{5/2}}{(1-wV)^2}(w-V),
\end{equation}
where $V=dZ/dT$, $\alpha$ is a characteristic of the plasma
clump, $w$ is a characteristic of the electromagnetic field
of the ambient medium and $k$ is essentially the curvature of
the central source given by
\begin{equation}\label{4.2}
k(T)=-\frac{2 GMz(z^2-3 a^2)}{(z^2+a^2)^3},
\end{equation}
while $z(T)$ is determined using the geodesic equation for the background
\begin{equation}\label{4.3}
\big(\frac{dz}{dT}\big)^2=\gamma^2-1+\frac{2GMz}{z^2+a^2}.
\end{equation}
\begin{figure}[t]
\centerline{\psfig{file=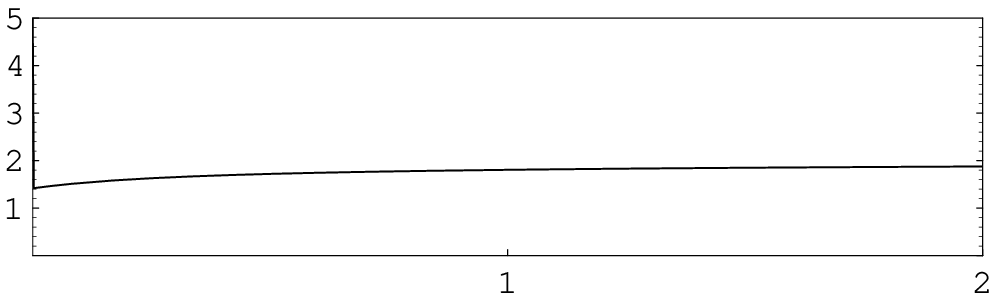, width=30pc}}
\centerline{\psfig{file=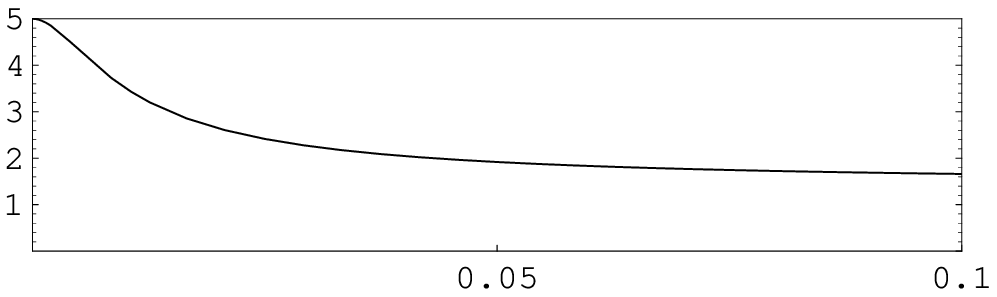, width=30pc}}
\caption{The top panel depicts the graph of the Lorentz factor $\Gamma$, 
given by $\Gamma(T):=(1-V(T)^2)^{-1/2}$,
versus time specified in hours. The bottom panel depicts the graph of $\Gamma$
versus time specified in seconds. For these graphs,
 $\widehat{a}=1$, $\gamma=1$,
$\widehat{\alpha}=10^{-7}$, $w=\sqrt{3}/2\approx .866$, $r_0/(GM)=z(0)=100$, 
$Z(0)=0$ 
and $V(0)=2 \sqrt{6}/5\approx 0.980$. We assume that $M=10 M_\odot$, hence
time is measured here in units of $GM=10\, GM_\odot\approx 5\times 10^{-5}\,$sec.
\label{fig:2}}
\end{figure}

The equation of motion~\eqref{4.1} has been derived for the clump
moving along the positive $Z$-axis. Recall that $z$ in equations~\eqref{4.2}
and~\eqref{4.3} is a substitute for the radial coordinate $r$. 
Therefore, let us note that equations~\eqref{4.1}--\eqref{4.3} are invariant
under the transformations $Z\to -Z$, $w\to -w$, $V\to -V$ and $z\to z$.
Hence our results hold for the clump moving along the negative $Z$-direction
as well.

It is useful
to put equations~\eqref{4.1}--\eqref{4.3} in dimensionless form.
To this end, we assume that all lengths are measured in units
of $GM$, which is one-half of the gravitational radius of the central source.
Defining dimensionless quantities $\widehat {k}$, $\widehat{\alpha}$
and $\widehat{a}$
as 
\begin{equation}\label{4.4}
\widehat {k}=(GM)^2 k,\qquad \widehat{\alpha}=GM\alpha, \qquad
\widehat{a}=\frac{a}{GM}
\end{equation}
we recover equations~\eqref{4.1}--\eqref{4.3} in dimensionless form once
we let $k\to\widehat {k}$, $\alpha\to \widehat{\alpha}$, $a\to \widehat{a}$ and $GM\to 1$.

The equations of motion in dimensionless form are 
equivalent to the dynamical system
\begin{eqnarray}\label{dysys}
\nonumber \frac{dz}{dT}&=& 
\Big(\gamma^2-1+\frac{2z}{z^2+\widehat{a}^2}\Big)^{1/2},\\
\nonumber \frac{dZ}{dT}&=& V,\\
\frac{dV}{dT}&=& \widehat{\alpha} \frac{(1-V^2)^{5/2}}{(1-wV)^2}(w-V)+
\frac{2 z(z^2-3 \widehat{a}^2)}{(z^2+\widehat{a}^2)^3}(1-2 V^2) Z,
\end{eqnarray} 
where we assume that $\gamma\ge 1$.

We have proved the following result: If $\widehat{\alpha}>0$,
$0\le w<1$,  $z(0)>\sqrt{3}\, \widehat {a}$, and $0< V(0)<1$, then
$\lim_{T\to\infty}V(T)=w$. That is, as time increases
the clump speed approaches the drift speed of the ambient medium.

Equation~\eqref{4.1} and the system~\eqref{dysys}
show how the tidal equation can in principle
be generalized to include nongravitational forces. On the other hand, it is
necessary to emphasize the qualitative significance of equation~\eqref{4.1}, 
since its
gravitational part is valid only within a distance $\mathcal{R}$
 of the fiducial
particle and its electromagnetic part is based on a rather simple model. 
If the
tidal part is completely ignored, then the Lorentz force would lead to a
monotonic deceleration of an initially ultrarelativistic clump toward $ w$.
Numerical experiments based on the system~\eqref{dysys}
demonstrate that the tidal part
initially dominates, causing a very rapid drop in $\Gamma$ toward $\sqrt{2}$, and
this occurs mostly within the domain of validity of the tidal term, but then
the electromagnetic term takes over and $\Gamma$ slowly tends toward $( 1 - w^2 )^
{-1/2}$.
Figure~\ref{fig:2} depicts a
profile for the graph of $\Gamma$ versus time, where the
parameter values and initial conditions are chosen to be typical
for a microquasar with $M=10 M_\odot$. According to our model~\eqref{dysys}
with  electrical
drift speed corresponding to $\Gamma=2$ and with Lorentz
acceleration coefficient $\widehat{\alpha}=10^{-7}$, an initial 
Lorentz factor of $\Gamma=5$ decreases to $\Gamma\approx 1.5$ 
in about $0.25\,\mbox{sec}$
and then very slowly increases to within 1\% of $\Gamma=2$ after
48 hours.  More generally, for $\widehat{\alpha}<10^{-2}$ 
the  timescale for the initial rapid decrease in $\Gamma$ is about
$0.1\,\mbox{sec}$.  The corresponding total relaxation time
to $\Gamma=2$ is approximately
$(50\widehat{\alpha})^{-1}\,\mbox{sec}$, so that for $\widehat \alpha$
ranging from $10^{-3}$ to $10^{-11}$, it ranges from 
$20\,\mbox{seconds}$ to $64 \,\mbox{years}$. 
For a quasar with $M=10^8 M_\odot$, the timescale for the initial
rapid decrease in $\Gamma$ would be about two weeks. 
\begin{figure}[t]
\centerline{\psfig{file=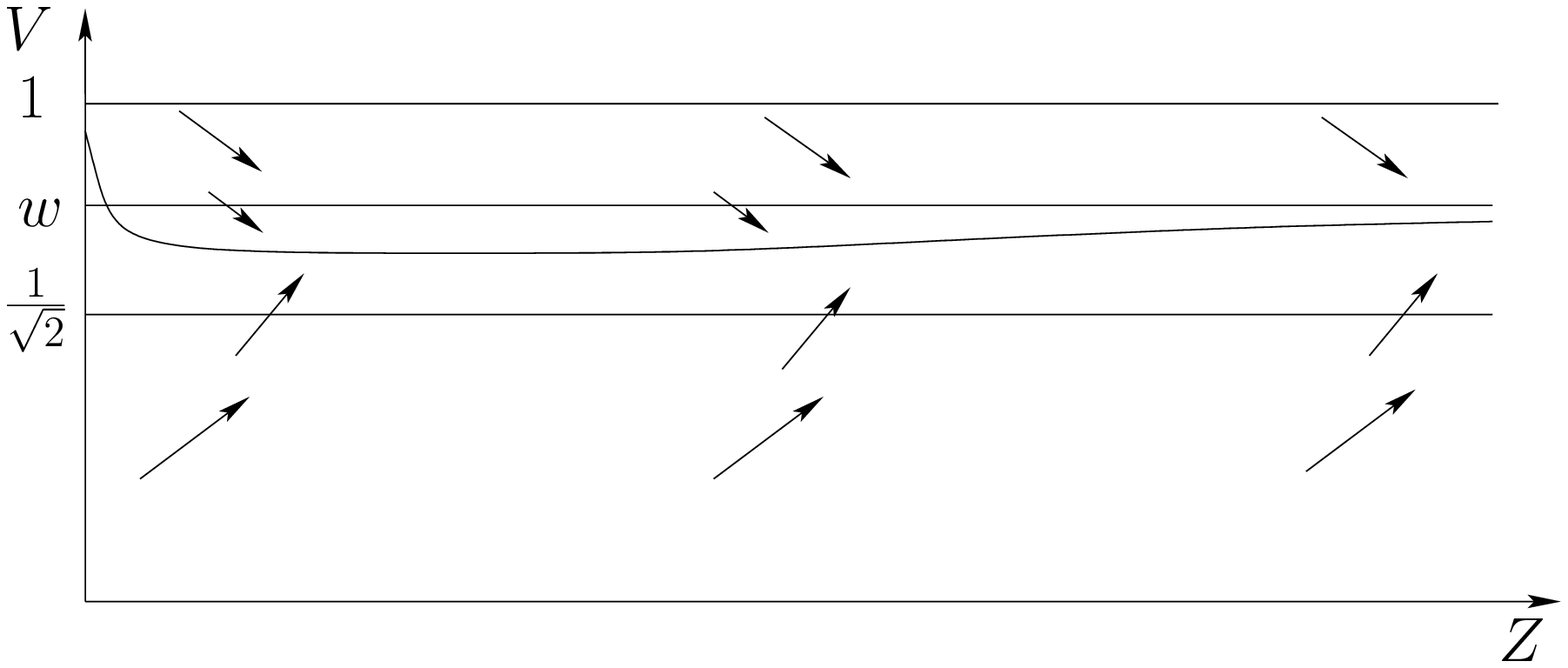,width=30pc}}
\centerline{\psfig{file=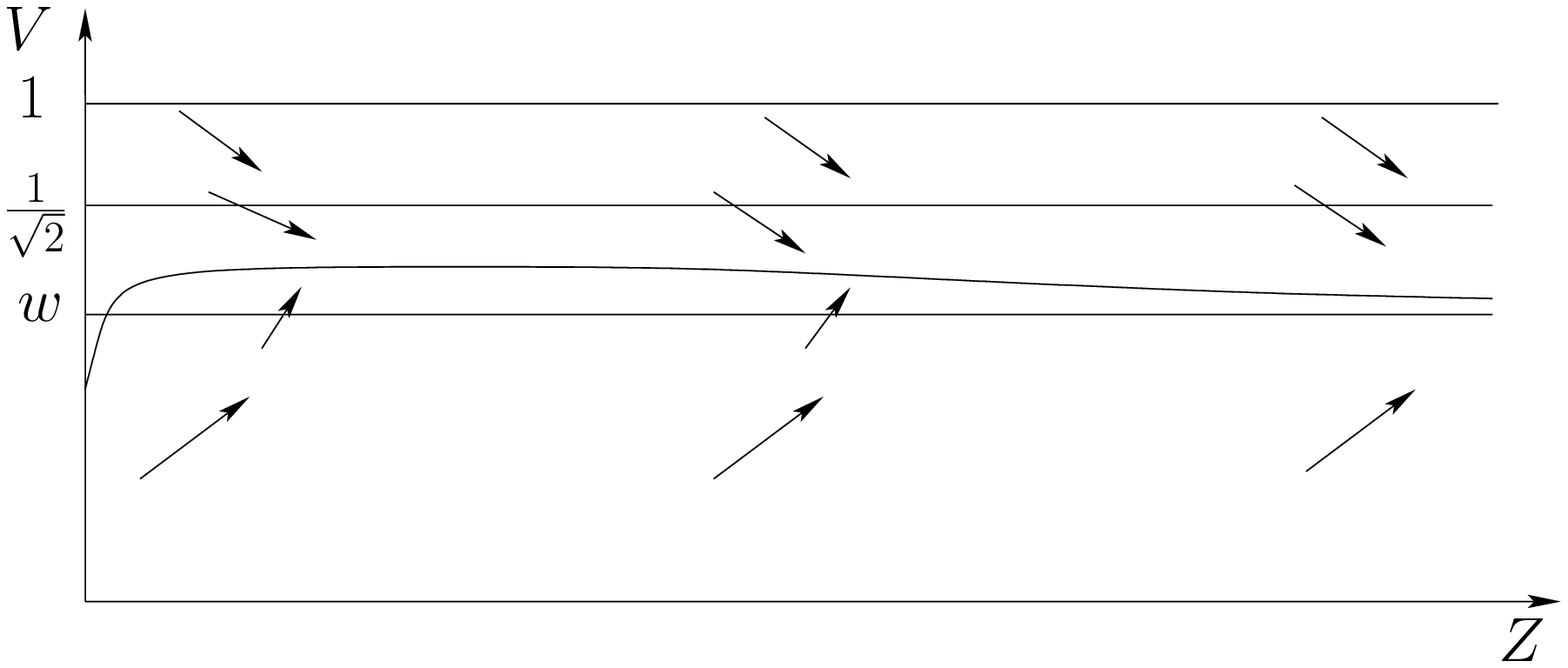,width=30pc}}
\caption{Schematic representation of direction fields and  trajectories for
the dynamical system~\eqref{dysys3}.
\label{fig:3}}
\end{figure}

To see that our result holds in general, let us note that
$dz/dT>0$ for $T>0$. Thus, $z$ is an increasing function of 
$T$, which is defined for $0\le T<\infty$. By viewing $z$ as 
a new ``time'' in the dynamical system~\eqref{dysys},
it suffices to consider the system
\begin{eqnarray}\label{dysys2}
\nonumber \frac{dZ}{dz}&=& \frac{1}{R(z)} V,\\
\frac{dV}{dz}&=& \frac{1}{R(z)}\Big(\widehat{\alpha} \frac{(1-V^2)^{5/2}}{(1-wV)^2}(w-V)+
\frac{2 z(z^2-3 \widehat {a\,}^2)}{(z^2+\widehat{ a\,}^2)^3}(1-2 V^2) Z\Big),
\end{eqnarray}
where 
\[R(z):=\Big(\gamma^2-1+\frac{2z}{z^2+\widehat{ a\,}^2}\Big)^{1/2}.\]
In this system $dZ/dz>0$. Hence, as before,
it suffices to consider the scalar differential equation
\begin{equation}\label{dysys3}
\frac{dV}{dZ}=\widehat{\alpha} \frac{(1-V^2)^{5/2}}{(1-wV)^2 V}(w-V)
+ 2 Z\frac{ S(Z)(S(Z)^2-3 \widehat{ a\,}^2)(1-2 V^2)}{(S(Z)^2+\widehat{ a\,}^2)^3 V},
\end{equation}
where $S$ is the function such that  $S(Z(z))=z$. 

Direction fields of $dV/dZ$ in the $(Z,V)$-plane are depicted
in Figure~\ref{fig:3} for the two cases $w>1/\sqrt{2}$ and 
$w<1/\sqrt{2}$. Graphs of typical solutions are
also drawn.   Since $V$ stays bounded and the curvature $k(T)$
approaches zero very fast,  it can be shown that the
second term on the right-hand side of equation~\eqref{dysys3} rapidly 
approaches zero as $T\to \infty$. Also, with this term set to zero, the
dynamical system~\eqref{dysys3} has an asymptotically stable steady
state at $V=w$. For large $Z$, the first term is dominant; therefore, 
the limiting value of the solution is $w$.

\section{Discussion}\label{sec:5}
This paper is devoted to a simple model of the evolution of the
speed of a clump in a relativistic flow once it emerges from the
environment immediately surrounding the central source. Combined with
a realistic treatment of plasma effects, our theoretical approach should
be of interest in the study of astrophysical jets.
If the
motion of the clump is force-free, then it follows from the repeated application
of the
generalized Jacobi equation that the clump speed tends to 
$V_T =1/\sqrt{2}$ corresponding to $ \Gamma = \sqrt{2}$.
On the other hand, if
the force-free condition is not satisfied, the clump speed 
tends to $V_L = w$, where $w$ is the electrical drift speed.
In fact, if the tidal component initially dominates, as would be the
case for most microquasars, then starting with an initial speed almost
equal to unity,
the tidal force is responsible for the initial rapid decrease of the velocity
toward $1/\sqrt{2}$, but then the Lorentz term takes over and the
velocity slowly approaches $w$. On the other hand, if the
Lorentz term is dominant, as would be the case for most quasars, there
is in effect a relatively slow decrease of the velocity toward $w$.

\end{document}